\documentstyle[sprocl]{article}

\input epsf.tex

\def\be{\begin{equation}}
\def\ee{\end{equation}}
\def\bea{\begin{eqnarray}}
\def\eea{\end{eqnarray}}
\begin{document}

\title{PROTON SPIN STRUCTURE AND \\
THE LOW-ENERGY $pp \rightarrow pp \eta'$ REACTION}

\author{STEVEN D. BASS}

\address{Physik Department T39, Technische Universit\"at M\"unchen,\\
D-85747 Garching, Germany}
%\\E-mail: Steven.Bass@cern.ch}


\maketitle\abstracts{ 
Gluonic degrees of freedom induce a contact term in the effective chiral 
Lagrangian description of the low-energy 
$pp \rightarrow pp \eta'$ reaction.
The strength of this contact term is, in part, related to 
the amount of spin carried by polarised gluons in a polarised proton.}

Polarised deep inelastic scattering and $\eta'$ physics provide complementary 
windows on dynamics induced by the axial $U_A(1)$ anomaly in QCD 
\cite{bass99a}. 
The flavour-singlet Goldberger-Treiman relation \cite{venez} 
relates the flavour-singlet axial-charge $g_A^{(0)}$ measured in 
polarised deep inelastic scattering to the 
$\eta'$--nucleon coupling constant $g_{\eta' NN}$.
The large mass of the $\eta'$ and the small value of $g_A^{(0)}$ 
\begin{equation}
\left. g^{(0)}_A \right|_{\rm pDIS} = 0.2 - 0.35
\end{equation}
extracted from deep inelastic scattering point to substantial violations of 
the OZI rule in the flavour-singlet $J^P=1^+$ channel.

Working in the chiral limit the flavour-singlet Goldberger-Treiman relation
reads
\begin{equation}
m g_A^{(0)} = \sqrt{3 \over 2} F_0 \biggl( g_{\eta' NN} - g_{QNN} \biggr) 
\end{equation}
where
$g_{\eta' NN}$ is the $\eta'$--nucleon coupling constant and $g_{QNN}$ 
is an OZI violating coupling which measures the coupling of 
the topological charge density $Q = {\alpha_s \over 4 \pi} G {\tilde G}$
to the nucleon.
In Eq.(2) $m$ is the nucleon mass and $F_0$ ($\sim 0.1$GeV) renormalises 
the flavour-singlet decay constant.
The coupling constant $g_{QNN}$ is, in part,  related \cite{venez} 
to the amount of spin carried by polarised gluons in a polarised proton.
A large positive $g_{QNN} \sim 2.45$
is one possible explanation of the small value of $g_A^{(0)}|_{\rm pDIS}$.

It is important to look for other observables which are sensitive to 
$g_{QNN}$.  
OZI violation in the $\eta'$--nucleon system is a probe of the role 
of gluons in dynamical chiral symmetry breaking in low-energy QCD.

Working with the $U_A(1)$--extended chiral Lagrangian for low-energy QCD 
we find a gluon-induced contact interaction in 
the $pp \rightarrow pp \eta'$ reaction close to threshold \cite{bass99b}:
\begin{equation}
{\cal L}_{\rm contact} =
         - {i \over F_0^2} \ g_{QNN} \ {\tilde m}_{\eta'}^2 \
           {\cal C} \
           \eta' \ 
           \biggl( {\bar p} \gamma_5 p \biggr)  \  \biggl( {\bar p} p \biggr)
\end{equation}
Here ${\tilde m}_{\eta_0}$ is the gluonic contribution to the $\eta'$ 
mass and ${\cal C}$ is a second OZI violating coupling which also features 
in $\eta'N$ scattering.
The physical interpretation of the contact term (3) 
is a ``short distance'' ($\sim 0.2$fm) interaction 
where glue is excited in the interaction region of
the proton-proton collision and 
then evolves to become an $\eta'$ in the final state.
This gluonic 
contribution to the cross-section for $pp \rightarrow pp \eta'$ 
is extra to the contributions associated with meson exchange models.

What is the phenomenology of this gluonic interaction ?

Since glue is flavour-blind the contact interaction (3) has the same 
size in both 
the $pp \rightarrow pp \eta'$ and $pn \rightarrow pn \eta'$ reactions.

CELSIUS \cite{celsius} have measured the ratio
$R_{\eta} 
 = \sigma (pn \rightarrow pn \eta ) / \sigma (pp \rightarrow pp \eta )$
for quasifree $\eta$ 
production from a deuteron target up to 100 MeV above threshold.
They observed that $R_{\eta}$ is approximately energy-independent 
$\simeq 6.5$ over the whole energy range.
The value of this ratio signifies a strong 
isovector exchange 
contribution to the $\eta$ production mechanism \cite{celsius}.
This experiment should be repeated for $\eta'$ production.
The cross-section for $pp \rightarrow pp \eta'$ 
close to threshold has been measured at COSY \cite{cosy}.
The $pn \rightarrow pn \eta'$ measurement remains to be done.
The more important that the gluon-induced process (3) 
is in the 
$pp \rightarrow pp \eta'$ reaction the more one would expect 
$R_{\eta'} =
 \sigma (pn \rightarrow pn \eta' ) / \sigma (pp \rightarrow pp \eta' )$
to approach unity near threshold
after we correct for the final state 
interaction between the two outgoing nucleons.
(After we turn on the quark masses, the small $\eta-\eta'$ mixing angle 
 $\theta \simeq -18$ degrees means that the gluonic effect
 (3) should be considerably bigger in $\eta'$ production than $\eta$
 production.)

$\eta'$ phenomenology is characterised by large OZI violations.
It is natural to expect 
large gluonic effects in the $pp \rightarrow pp \eta'$ process. 
A measurement of $R_{\eta'}$ is urgently needed!

\end{document}